\newcommand*{\RELEASE}{}  

\documentclass[conference]{IEEEtran}
\IEEEoverridecommandlockouts
\usepackage{cite}
\usepackage{amsmath,amssymb,amsfonts}
\usepackage{graphicx}
\usepackage{textcomp}
\usepackage{xcolor}
\def\BibTeX{{\rm B\kern-.05em{\sc i\kern-.025em b}\kern-.08em
    T\kern-.1667em\lower.7ex\hbox{E}\kern-.125emX}}

\usepackage{stfloats} 
\usepackage{algorithm}
\usepackage{algpseudocode}  

\usepackage{balance}  

\usepackage{array}  
\usepackage{booktabs}  
\usepackage{multirow}  
\usepackage[bottom]{footmisc}  

\graphicspath{{fig/}{../fig/}}

\usepackage[normalem]{ulem}  

\usepackage{colortbl}

\usepackage[utf8]{inputenc}
\usepackage{paralist}  
\usepackage{enumitem}  
\usepackage{xspace}  
\usepackage{url}

\newcolumntype{!}{>{\global\let\currentrowstyle\relax}}
\newcolumntype{^}{>{\currentrowstyle}}
\newcommand{\rowstyle}[1]{\gdef\currentrowstyle{#1}%
  #1\ignorespaces
}

\definecolor{green(pigment)}{rgb}{0.0, 0.65, 0.31}

\ifdefined\RELEASE  
	\newcommand{\al}[1]{} 
	\newcommand{\add}[1]{#1}  
	\newcommand{\refine}[2]{#1}
	\newcommand{\del}[1]{}  

	\newcommand{\mkh}[1]{}
	\newcommand{\jgrm}[1]{}
\else
	\newcommand{\al}[1]{\textcolor{green(pigment)}{[AL: #1]}} 
	\newcommand{\add}[1]{\textcolor{blue}{#1}}  
	\newcommand{\refine}[2]{\textcolor{violet}{#1}\textcolor{teal}{[#2]}}
	\newcommand{\del}[1]{\textcolor{blue}{\sout{#1}}}  

	\newcommand{\mkh}[1]{\textcolor{brown}{(*** MK: #1 ***)}}
	\newcommand{\jgrm}[1]{\textcolor{brown}{\sout{#1}}}
\fi

\newcommand{\anm}{ANONYMOUS\xspace}
\ifdefined\ANONYMOUS
	\newcommand{\xms}{ANONYM_APP\xspace}  
	\newcommand{\urlXms}{\anm} 
	\newcommand{\bmark}{ANONYM_BMARK\xspace}  
	\newcommand{\urlBmark}{\anm}  
\else
	\newcommand{\xms}{xmeasures\xspace}  
	\newcommand{\urlXms}{\url{https://github.com/eXascaleInfolab/xmeasures}\xspace}
	\newcommand{\bmark}{Clubmark\xspace}  
	\newcommand{\urlBmark}{\url{https://github.com/eXascaleInfolab/clubmark}\xspace}  
\fi

\begin{document}


\title{Accuracy Evaluation of Overlapping and Multi-resolution Clustering Algorithms\\  
on Large Datasets
\thanks{
This project has received funding from the
\ifdefined\ANONYMOUS
\anm
\else
European Research Council (ERC) under the European Union’s Horizon 2020 research and innovation program (grant agreement 683253/GraphInt) and in part by the Swiss National Science Foundation under grant number CRSII2 147609.
\fi
}
}


\author{
\ifdefined\ANONYMOUS
\\
\anm\\
\\
\else
\IEEEauthorblockN{Artem Lutov, Mourad Khayati and Philippe Cudr{\'e}-Mauroux}
\IEEEauthorblockA{eXascale Infolab, University of Fribourg---Switzerland\\
Email: \{firstname.lastname\}@unifr.ch
}
\fi
}

\maketitle

\begin{abstract}
Performance of clustering algorithms is evaluated with the help of accuracy metrics. There is a great diversity of clustering algorithms, which are key components of many data analysis and exploration systems. However, there exist only few metrics for the accuracy measurement of overlapping and multi-resolution clustering algorithms on large datasets.
In this paper, we first discuss existing 
metrics, how they satisfy a set of formal constraints, and how they can be applied to specific cases.
Then, we propose several optimizations and extensions of these metrics. More specifically, we introduce a new indexing technique to reduce both the runtime and the memory complexity of the Mean F1 score evaluation. Our technique can be applied on large datasets 
and it is faster on a single CPU than state-of-the-art implementations running on high-performance servers. In addition, we propose several extensions of the discussed metrics to improve their effectiveness and satisfaction to formal constraints without affecting their efficiency. All the metrics discussed in this paper are implemented in C++ and are available for free as open-source packages that can be used either as stand-alone tools or as part of a benchmarking system to compare various clustering algorithms.
\end{abstract}



\section{Introduction}
\label{sec:intro}

Clustering is a key component of many data mining systems with numerous applications including statistical analysis and the exploration of physical, social, biological and informational systems. This diversity of potential applications spawned a wide variety of network (graph) clustering algorithms proposed in the literature. 
It also led to specialized clustering algorithms in particular domains. Hence, the need to find the most suitable and best performing clustering algorithms for a given task became more dire, 
and the evaluation of the resulting clustering through proper metrics more important. Moreover, as modern systems often operate on very large datasets (consisting of billions of items potentially), the computational properties of the evaluation metrics become more important. In particular, performance-related constraints rapidly emerge when sampling the original large datasets is not desirable or not possible for a given use-case.

Clustering quality metrics can formally be categorized into two types: intrinsic and extrinsic metrics.
Intrinsic quality metrics evaluate how the elements of each cluster are similar to each other and how they differ from elements in other clusters given a similarity metric such as modularity~\cite{Nwm04u} or conductance~\cite{Kan04}. 
Extrinsic quality metrics (also known as 
\emph{accuracy} metrics) evaluate instead how the clusters are similar to the \emph{ground-truth} (gold standard) clusters. In this paper, we focus on extrinsic metrics since they allow to identify clustering algorithms producing expected results, which is in practice often more useful than measuring the formation of optimal clusters by a given similarity metric.
More specifically, we evaluate the similarity (proximity) between two collections of overlapping clusters (unordered sets) of elements, where
\begin{inparaenum}[\itshape a\upshape)]
\item the collections have the same number of elements,
\item each element may be a member of multiple clusters and
\item each cluster may have several, non-mutual, best 
(in terms of the similarity value) matches in the other collection.
\end{inparaenum}

We call \emph{clustering} the set of clusters resulting from an algorithm. Clusterings can be categorized as non-overlapping (crisp clustering, hard partitioning), overlapping (soft, fuzzy clustering) or, in some cases, multi-resolution (including hierarchical). Multi-resolution clusterings are considered when there is a need to simultaneously compare multiple resolutions (hierarchy levels) of the results against the ground-ground, where each resolution contains non/overlapping clusters as discussed further in Section~\ref{sec:discussion}. Non-overlapping clusterings can be seen as a special case of overlapping clusterings.

A large number of accuracy metrics were proposed in the literature to measure 
the clustering quality~\cite{Amg09,Cln88,Dnn05,Rsb07,Yng13,Rlm13}. 
Evaluation frameworks~\cite{Chn14,Shr15,Yin16,Mzc15,Clb18} and surveys~\cite{Wan15,Hnb14,Hsc14,Sae18} were also introduced. 
Despite the large number of accuracy metrics proposed, very few metrics are applicable to overlapping clusters, 
causing many issues when evaluating such clusters (e.g., Adjusted Rand Index is used to evaluate overlapping clusters in~\cite{Mln11}, even though it is applicable to non-overlapping clusters only). 
Moreover, most of the quality metrics 
for overlapping clusters are not comparable to similar metrics for non-overlapping clusters (e.g., standard NMI~\cite{Dnn05} or modularity~\cite{Nwm04u} versus some overlapping NMI~\cite{Mdd11} or overlapping modularity~\cite{Zhn07,Ncs09,Lzr10,Chn15} implementations), which complicates the direct comparison of the respective clustering algorithms.

Therefore, further research is required to develop accuracy metrics that are applicable to overlapping (and multi-resolution) clusterings, that satisfy tight performance constraints and, ideally, that are compatible with the results of standard accuracy metrics used for non-overlapping clustering.
%
Finally,
producing a single, easy to interpret value for the final clustering is of importance also, in order to help the user pick the most suitable clustering for a particular use-case and for potentially several accuracy metrics. 
This issue has been tackled through the formal constraints, introduced for example in~\cite{Amg09,Rsb07}, and is further discussed 
in Section~\ref{sec:formconstr}.

To the best our knowledge, this is the first work discussing all state-of-the-art accuracy metrics applicable to overlapping clustering evaluation on large datasets (i.e., with more than $10^7$ elements).  
Being able to evaluate metrics on large datasets means---in our context---that the evaluation process should be at most:
\begin{inparaenum}[\itshape a\upshape)]
\item \emph{quadratic} in terms of runtime complexity and  
\item \emph{quasilinear} in terms of memory complexity
\end{inparaenum}
with the number of elements considered.
In addition, we also introduce in this paper a novel indexing technique to reduce the runtime complexity of the Mean F1 score (and a similar metric, NVD~\cite{Chn14}) evaluation from
$O(N \cdot (|C| + |C'|))$ to $O(N)$, where $N$ is the number of elements in the processed clusters, $|C|$ is the number of resulting clusters and $|C'|$ is the number of ground-truth clusters.
Finally, we propose extensions to the state-of-the-art accuracy metrics to satisfy more formal constraints (i.e., to improve their effectiveness) without sacrificing  their efficiency. Efficient C++ implementations of
\begin{inparaenum}[\itshape a\upshape)]
\item all discussed accuracy metrics and
\item their improved versions
\end{inparaenum}
are freely available online as open source utilities as listed in the sections devoted to each metric. All our accuracy metrics are also integrated into an open source benchmarking framework for clustering algorithms evaluation, \bmark\footnote{\urlBmark\label{ftn:bmsrc}}~\cite{Clb18}, besides being available as dedicated applications.

\section{Related Work}
\label{sec:relwork}

Related efforts can be categorized into three main groups, namely:
\begin{inparaenum}[\itshape a\upshape)]
\item accuracy metrics for overlapping clustering evaluation (satisfying the complexity constraints mentioned above),
\item frameworks providing efficient implementations for accuracy evaluation and
\item 
formal constraints for accuracy metrics (discussed in Section~\ref{sec:formconstr}).
\end{inparaenum}

\paragraph{Accuracy Metrics for Overlapping Clusters}
The \emph{Omega Index}~\cite{Cln88} is the first accuracy metric that was proposed for overlapping clustering evaluation. It belongs to the family of \emph{Pair Counting Based Metrics}. It is a fuzzy version of the Adjusted Rand Index (ARI)~\cite{Hbt85} and is identical to the Fuzzy Rand Index~\cite{Hlm09}. We describe the Omega Index in Section~\ref{sec:omega}.
\\  
Versions of \emph{Normalized Mutual Information (NMI)} suitable for overlapping clustering evaluation were introduced as Overlapping NMI (ONMI)\footnote{\url{https://github.com/eXascaleInfolab/OvpNMI}}~\cite{Mdd11} and Generalized NMI (GNMI)~\cite{Esv12} and belong to the family of \emph{Information Theory Based Metrics}. The authors of ONMI suggested to extend Mutual Information with  approximations (introduced in~\cite{Lcn09}) to find the best matches for each cluster of a pair of overlapping clusterings. This approach allows to compare overlapping clusters, but unlike GNMI we introduce in Section~\ref{sec:gnmi}, it yields values that are incompatible with standard NMI~\cite{Dnn05} results.
\\
The \emph{Average F1 score} is introduced in~\cite{Yng13,Prat14} and a similar metric, NVD, is introduced in~\cite{Chn14}. The Average F1 score belongs to the family of \emph{Cluster Matching Based Metrics} and is described in Section~\ref{sec:mf1}.

\paragraph{Accuracy Measurement Frameworks}
A toolkit\footnote{\url{https://github.com/chenmingming/ParallelComMetric}} for the parallel measurement of the quality of non-overlapping clusterings on both distributed and shared memory machines is introduced in~\cite{Chn14}. This toolkit performs the evaluation of several accuracy metrics (Average F1 score, NMI, ARI and JI) as well as some intrinsic quality metrics, and provides highly optimized parallel implementations of these metrics in C++ leveraging MPI (the Message Passing Interface) and Pthreads (POSIX Threads). Among its 
accuracy metrics, 
only Average F1 score is applicable to overlapping clusterings.
\\  
\emph{WebOCD}~\cite{Shr15} is an open-source RESTful web framework for the development, evaluation and analysis of \emph{overlapping} community detection (clustering) algorithms. It comprises several baseline algorithms, evaluation metrics and 
preprocessing utilities. 
However, since WebOCD (including all its accuracy metrics) is implemented in pure Java as a monolithic framework,
many existing implementations of evaluation metrics cannot be easily integrated into WebOCD without either being reimplemented in Java or modifying the framework architecture. 
A reimplementation of existing metrics is not always possible
without a significant performance drop
(especially when linking native, high-performance libraries such as Intel TBB, STL or Boost)
and time investment.
\\
\emph{CoDAR}~\cite{Yin16} is a framework for 
community detection algorithm evaluation and recommendation providing user-friendly interfaces and visualizations.
Based on this framework, the authors also introduced a study of \emph{non-overlapping} community detection algorithms on \emph{unweighed undirected} networks~\cite{Wan15}.
Unfortunately, the framework URL provided in the paper refers to a forbidden page, i.e. the implementation is not available to the public anymore.
\\
\emph{Circulo}~\cite{Mzc15} is a framework for community detection algorithms evaluation. It executes the algorithms on preliminary uploaded input networks and then evaluates the results using several accuracy metrics and multiple intrinsic metrics.

\section{Formal Constraints on Clustering Evaluation Metrics}
\label{sec:formconstr}

\begin{figure}[tbp]
\begin{minipage}[t]{0.48\textwidth}
\centering
\includegraphics[scale=0.2]{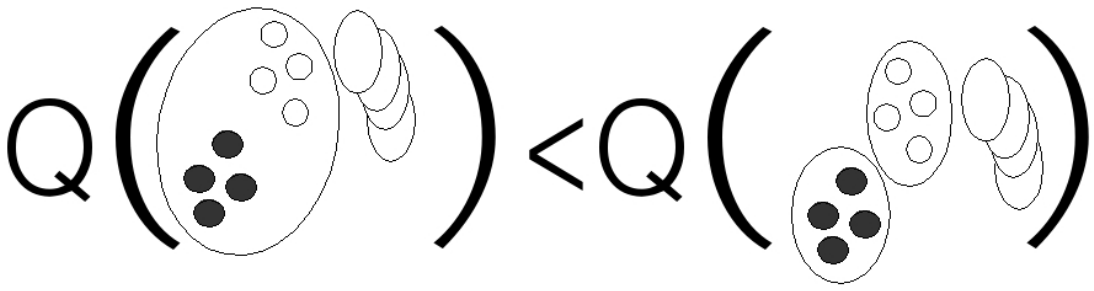}
\caption{Homogeneity. Clusters should not mix elements belonging to different categories.}
\label{fig:homogeneity}
\end{minipage}
\hfill
\vspace{8pt}
\begin{minipage}[t]{0.48\textwidth}
\centering
\includegraphics[scale=0.2]{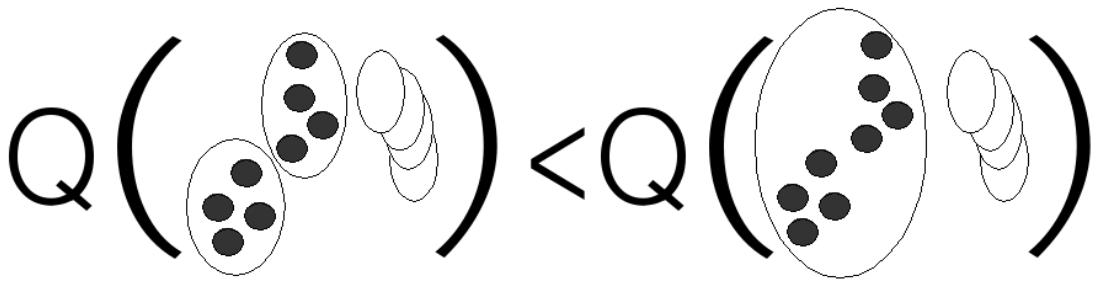}
\caption{Completeness. Elements belonging to the same category should be clustered together.}
\label{fig:completeness}
\end{minipage}
\hfill
\vspace{8pt}
\begin{minipage}[t]{0.48\textwidth}
\centering
\includegraphics[scale=0.2]{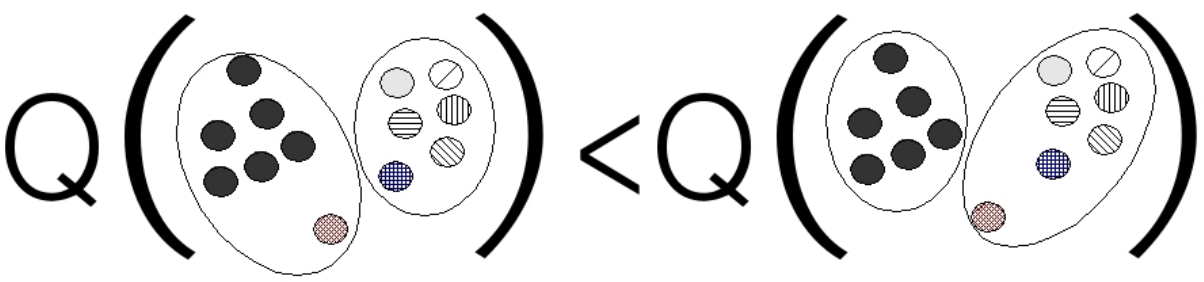}
\caption{Rag Bag. Elements with low relevance to the categories (e.g., noise) should be preferably assigned to the less homogeneous clusters (macro-scale, low-resolution, coarse-grained or top-level clusters in a hierarchy).}
\label{fig:ragbag}
\end{minipage}
\hfill
\vspace{8pt}
\begin{minipage}[t]{0.48\textwidth}
\centering
\includegraphics[scale=0.2]{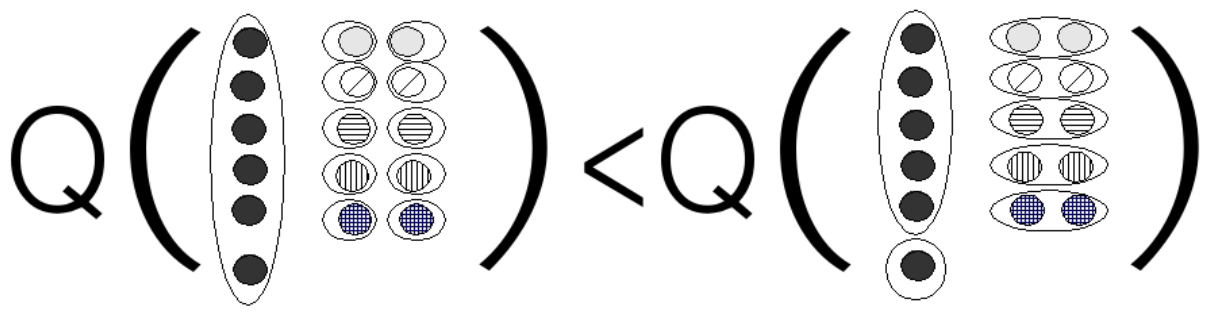}
\caption{Size vs Quality (Micro Weighting). A small assignment error in a large cluster is preferable to a large number of similar errors in small clusters.}
\label{fig:szquality}
\end{minipage}
\vspace{-6pt}
\end{figure}

Four formal constraints for the accuracy metrics were introduced in~\cite{Amg09} and shed light on which aspects of the quality of a clustering are captured by different metrics. 
Two of these constraints (Homogeneity, see Fig.~\ref{fig:homogeneity} and Completeness, see Fig.~\ref{fig:completeness}) were originally proposed in~\cite{Rsb07} while the other two (Rag Bag, Fig.~\ref{fig:ragbag} and Size vs Quantity, Fig.~\ref{fig:szquality}) were newly introduced. 
Besides been intuitive, these four constraints were developed to be formally provable and to clarify the limitations of each metric. We list these constraints in the way they were presented in~\cite{Amg09} and later use them to discuss various accuracy metrics. Each constraint is written as an inequality applied to a pair of clusterings, where a quality metric Q (some accuracy metric in our case) from the clustering on the right-hand side is assumed to be better than the one from the left-hand side.
Ground-truth clusters are called \emph{categories} for short in the following.


\section{Accuracy Metrics for Overlapping Clustering}
\label{sec:accumtr}

\begin{figure}[htbp]
\begin{minipage}[t]{0.24\textwidth}
\centering
\includegraphics[scale=0.25]{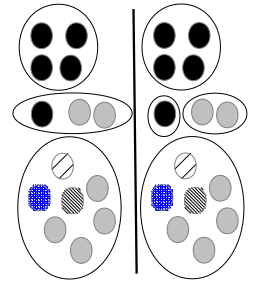}
\caption{Homogeneity sample.}
\label{fig:smphomogeneity}
\end{minipage}
\begin{minipage}[t]{0.24\textwidth}
\centering
\includegraphics[scale=0.25]{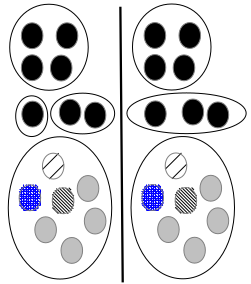}
\caption{Completeness sample.}
\label{fig:smpcompleteness}
\hfill
\end{minipage}
\begin{minipage}[t]{0.24\textwidth}
\centering
\includegraphics[scale=0.25]{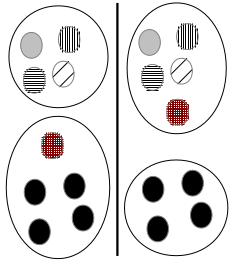}
\caption{Rag Bag sample.}
\label{fig:smpragbag}
\end{minipage}
\begin{minipage}[t]{0.24\textwidth}
\centering
\includegraphics[scale=0.25]{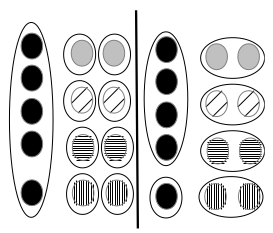}
\caption{Size vs Quality sample.}
\label{fig:smpszquality}
\end{minipage}
\vspace{-6pt}
\end{figure}

Accuracy metrics indicate how much
one clustering (i.e., set of clusters)  
is similar to another (ground-truth) clustering.
For each presented metric, we first give its original definition before proposing our extensions and optimizations. Then, we empirically evaluate the aforementioned four formal constraints on samples from~\cite{Amg09} and given in Fig.~\ref{fig:smphomogeneity}-\ref{fig:smpszquality} denoting the left clustering as \emph{Low} and the right clustering as \emph{High} for each sample. 
The constraints that are satisfied are marked in italics
in the results tables for each discussed metric in the respective subsection. 
Cluster elements belonging to the same category (ground-truth cluster) in these figures are colored with the same color and texture, while the formed clusters (results) 
are shown with oval shapes.

\subsection{Omega Index}  
\label{sec:omega}

\subsubsection{Preliminaries}

The Omega Index~\cite{Cln88} is an ARI~\cite{Hbt85} generalization applicable to overlapping clusters. It is based on counting 
the number of pairs of elements occurring in exactly the same number of clusters as 
in the number of categories and adjusted to the expected 
number of such pairs.
Formally, given the ground-truth clustering $C'$ consisting of categories $c'_i \in C'$ and formed clusters $c_i \in C$:
\begin{equation}
Omega(C',C) = \frac{Obs(C',C) - Exp(C',C)}{1 - Exp(C',C)}.
\label{eq:omega}
\end{equation}
The observed agreement is:
\begin{equation}
Obs(C',C) = \sum_{j = 0}^{\min(J', J)}{A_j / P},
\label{eq:omegaobs}
\end{equation}
where $J'$ ($J$) is the maximal number of categories (clusters) in which a pair of elements occurred, 
$A_j$ is the number of pairs of elements occurring in
exactly $j$ categories and exactly $j$ clusters,
and $P = N \cdot (N - 1) / 2$ is the total number of pairs given a total of $N$ elements (nodes of the network being clustered).\\
The expected agreement is:
\begin{equation}
Exp(C',C) = \sum_{j = 0}^{\min(J', J)}{P'_j P_j / P^2},
\label{eq:omegaexp}
\end{equation}
where $P'_j$ ($P_j$) is the total number of pairs of elements assigned to exactly 
$j$ 
categories (clusters).  

\subsubsection{Proposed Extension (Soft Omega Index)}

The Omega Index evaluates overlapping clusterings by counting the number of pairs of elements present in exactly the same number of clusters as in the number of categories, which does not take into account pairs present in slightly different number of clusters.
We propose to fix this issue by normalizing smaller number of occurrences of each pair of elements in all clusters of one clustering by the larger number of occurrences in another clustering as outlined on line~\ref{aln:oagrnorm} of Algorithm~\ref{alg:omegasoft}. The input data consists of two clusterings ($grs, cls$), and the $rels$ hashmap relating the clusters to their elements (nodes) for each clustering. 
The updated computation of the observed agreement of pairs requires also to correct the expected agreement, which is performed on line~\ref{aln:eagrcorr}.
\begin{algorithm}[htbp]  
\caption{Soft Omega Index}
\label{alg:omegasoft}
\begin{algorithmic}[1]  
\Procedure{omegaSoft}{$rels, grs, cls$} 
\State $nobs$ = 0  \Comment{Observed number}
\State $ngs$ = vector(size($grs$))  \Comment{Ranked pairs counts for grs}
\State $ncs$ = vector(size($cls$))  \Comment{Ranked pairs counts for cls}
\For{$ir$ in range(begin($rels$), end($rels$))}
	\For{$jr$ in range(++$ir$, end($rels$))}
		\State $gn$ = mutualnum(grs($ir$), grs($jr$))
		\State $cn$ = mutualnum(cls($ir$), cls($jr$))
		\State $nobs$ += 1 if $gn$ == $cn$ else min($gn, cn$) / max($gn, cn$) \label{aln:oagrnorm}
		\State ++$ngs$[$gn$]; ++$ncs$[$cn$]
	\EndFor
\EndFor
\State $nexp$ = 0; $szmin$ = min(size($ngs$), size($ncs$))
\For{$i$ in range($szmin$)}
	\State $nexp$ += $ngs$[$i$] $\cdot\; ncs$[$i$]
\EndFor
\State $rns$ = $ngs$ if size($ngs$) $>\; szmin$ else $ncs$
\For{$i$ in range($szmin$, size($rns$))}
	\State $nexp$ += $rns$[$i$] \label{aln:eagrcorr}
\EndFor
\State $nps$ = size(nodes($rels$)); $nps\; *$= ($nps$ - 1) / 2
\State $nexp$ /= $nps$
\State \Return ($nobs - nexp$) / ($nps - nexp$)
\EndProcedure
\end{algorithmic}  
\end{algorithm}
\\  
Thus, \emph{OmegaSoft} has the same definition as Eq.~\ref{eq:omega}, except the observed agreement number is evaluated as:
\begin{equation}
Obs_{soft}(C',C) = \sum_{j = 0}^{\max(J', J)}{Anorm_j / P}\,,
\label{eq:omegsfaobs}
\end{equation}
where $Anorm_j$ is the number of pairs of elements occurring in 
exactly $j'$ and $j$ clusters of the clusterings and being weighted by $\min(j', j) / \max(j', j)$.\\
The expected agreement is:
\begin{equation}
Exp_{soft}(C',C) = (\sum_{j = 0}^{Jmin}{P'_j P_j} + \sum_{j = Jmin + 1}^{\max(J', J)}Prem_j) / P^2,
\label{eq:omegsfaexp}
\end{equation}
where $Jmin = \min(J', J)$, $Prem_j = P_j$ if $\min(J', J) = J'$ 
and $Prem_j = P'_j$ otherwise.
\\
Note that for non-overlapping clusterings (i.e., when the membership in clusters equals to 1 $\implies J' = J = 1$), the Soft Omega Index is equivalent to the original Omega Index, which is equivalent to ARI.

\subsubsection{Evaluation and Constraints Matching}

A counterexample outlining the issue of the Omega Index when discarding partially matching pairs of elements is shown in Table~\ref{tbl:oissue}, where each of the four categories (C1'-C4') consists of 3 elements and the total number of elements is 4 (\#1-\#4). The first clustering algorithm has a \emph{Low} accuracy and discovers only two clusters as shown in Table~\ref{tbl:oissue}. The second clustering algorithm (\emph{High}) performs much better discovering all four clusters but not all elements of the respective categories as shown in Table~\ref{tbl:oissue}. The original Omega Index fails to discriminate these cases yielding 0 for both cases, whereas the Soft Omega Index clearly differentiates the more accurate solution.
\begin{table}[htbp]
\centering
\caption{Accuracy evaluation of \textit{Low} and \textit{High} vs \textit{Ground-truth} clusterings by Omega Index and Soft Omega Index.}
\label{tbl:oissue}
\begin{tabular}{@{}!l^c^c@{}}
\toprule
\multirow{2}{*}{\textbf{Metrics \,\textbackslash\, Clusterings}} & \multicolumn{2}{c}{\begin{tabular}[c]{@{}c@{}}\textbf{Ground-truth}\\C1':\quad 1 2 3\\C2':\quad 2 3 4\\C3':\quad 3 4 1\\C4':\quad 4 1 2\end{tabular}}                                                                   \\ \cmidrule(l){2-3} 
                                                             & \multicolumn{1}{c|}{\begin{tabular}[c]{@{}c@{}}\textbf{Low}\\\\C1:\quad 1 2\\C2:\quad 3 4\\\\\end{tabular}} & \multicolumn{1}{c}{\begin{tabular}[c]{@{}c@{}}\textbf{High}\\C1:\quad 1 2\\C2:\quad 2 3\\C3:\quad 3 4\\C4:\quad 4 1\end{tabular}} \\ \midrule
Omega Index                                                  & 0                                                                      & 0                                                                                  \\
\rowstyle{\itshape}
Soft Omega Index                                             & 0                                                                      & 0.33                                                                               \\ \bottomrule
\end{tabular}
\end{table}
\\
The empirical satisfaction of the formal constraints for both versions of Omega Index is given in Table~\ref{tbl:constromega} and discussed below in Section~\ref{sec:discussion}. 
The computational complexity 
$O(N^2)$ and the memory complexity is $O(|C| + |C'|)\approx$\footnote{On average, the number of clusters in most real-world networks consisting of $N$ nodes is $\sqrt{N}$\label{ftn:clsnum}} $O(\sqrt{N})$ for both implementations, where $N$ is the number of elements in the clustering. Implementations of both the original and Soft Omega Index are provided in the open source \emph{\xms}\footnote{\urlXms\label{ftn:xmsrc}} utility and are available for free.
\begin{table}[htbp]
\centering
\caption{Formal Constraints for Soft and original Omega Index.}
\label{tbl:constromega}
\setlength\tabcolsep{2pt}  
\begin{tabular}{@{}l|cc|cc|cc|cc@{}}
\toprule
\multicolumn{1}{r|}{\textbf{Clusterings}} & \multicolumn{2}{c|}{\textbf{Homogen.}} & \multicolumn{2}{c|}{\textbf{Complet.}} & \multicolumn{2}{c|}{\textbf{RagBag}} & \multicolumn{2}{c}{\textbf{SzQual.}} \\
\multicolumn{1}{l|}{\textbf{Metrics}}                          & low                & high              & low                & high              & low              & high              & low               & high              \\ \midrule
{[}Soft{]} Omega Index                    & \textit{0.247}              & \textit{0.282}             & \textit{0.244}              & \textit{0.311}             & 0.4              & 0.4               & 0.804             & 0.804             \\ \bottomrule
\end{tabular}
\end{table}

\subsection{Mean F1 Score}  
\label{sec:mf1}

\subsubsection{Preliminaries}

The \emph{Average F1 score (F1a)} is a commonly used metric to measure the accuracy of clustering algorithms~\cite{Yng13,Prat14,Yan15}. F1a is defined as the average of the weighted F1 scores~\cite{Rjsb79} of
\begin{inparaenum}[\itshape a\upshape)]
\item the \emph{best matching} ground-truth clusters to the formed clusters and 
\item the \emph{best matching} formed clusters to the ground-truth clusters. 
\end{inparaenum}
Formally, given the ground-truth clustering $C'$ consisting of clusters $c'_i \in C'$ (called categories) and clusters $c_i \in C$ formed by the evaluating clustering algorithm:
\begin{equation}
F1a(C',C) = \frac{1}{2}(F_{C',C} + F_{C,C'}),
\label{eq:f1a}
\end{equation}
where
\begin{multline}
F_{X,Y} = \frac{1}{|X|}\sum_{x_i \in X} F1(x_i, g(x_i, Y)),\\
g(x, Y) = \{\texttt{argmax}_y\; F1(x, y)\; |\; y \in Y\},\quad
\label{eq:fxy}
\end{multline}
where $F1(x,y)$ is the F1 score of the respective clusters. 

\subsubsection{Proposed Extensions (F1h and F1p)}
The F1a definition yields \emph{non-indicative} values of $F1a \in [0, 0.5]$ when evaluating a large number of clusters. In particular, for clusters formed 
by taking all possible combinations of the nodes,  
$F1a>0.5\, (F1_{C',C} = 1$ since for each category there exists the exactly matching cluster, $F1_{C,C'} \rightarrow 0$ since majority of clusters have low similarity to the categories$)$.
To address this issue, 
we suggest to use the harmonic mean instead of the arithmetic mean (average). We introduce the \emph{harmonic F1 score (F1h)} as:
\begin{equation}
F1h(C',C) = \frac{2 F_{C',C} F_{C,C'}}{F_{C',C} + F_{C,C'}}.
\label{eq:f1h}
\end{equation}
$F1h \le F1a$ since the harmonic mean cannot be larger than the arithmetic mean. \add{\refine{In our case}{For the F1h}, $F_{C',C}$ can be interpreted as a recall and $F_{C,C'}$ as a precision of the evaluating clustering $C$ and the ground-truth clustering $C'$.}
\\
$F1h$ is more indicative than $F1a$ but both measures do not satisfy the Homogeneity constraint, as they penalize local best matches too severely. We propose to evaluate the probability of the local best matches rather than the F1 score to address the outlined issue. 
Our new metric $F1p$ is the harmonic mean (i.e. F1 measure) of the average over each clustering of the best local probabilities for each cluster. $F1p$ corresponds to the expected probability of the best match of the clusterings unlike $F1h$, which corresponds to the expected worst-case of the best match in the clusterings. Formally, $F1p$ is evaluated similarly to $F1h$, except that the local matching function $pprop$ given in Eq.~\ref{eq:pprob} replaces $f1$ given in Eq.~\ref{eq:f1}.  
\begin{equation}
pprob(m,c',c) = m/|c'| * m/|c|=\frac{m^2}{|c'| * |c|}\;, 
\label{eq:pprob}
\end{equation}
\begin{equation}
f1(m,c',c) = 2 \frac{m/|c'| * m/|c|}{m/|c'| + m/|c|}
= 2 \frac{m}{|c'| + |c|}\;, 
\label{eq:f1}
\end{equation}
where $m$ is the \emph{contribution} of matched elements between the cluster $c$ and category $c'$. The notations of contribution and $|x|$ (\texttt{size} of the cluster $x$) vary for the overlapping and multi-resolution clusterings and are discussed in Section~\ref{sec:discussion}. For multi-resolution and non-overlapping clusterings, $|x|$ is simply equal to the number of elements in the cluster $x$, and the contribution of each element is equal to 1. For overlapping clusterings, $|x|$ is equal to the total contribution of elements in cluster $x$, where each element $x_i$ contributes the value $1$/\texttt{shares}($x_i$) given that $x_i$ is a member of the number \texttt{shares}($x_i$) of (overlapping) clusters.

\subsubsection{Optimizations (Efficient Indexing) for the F1 Metric Family}
We propose an efficient indexing technique to reduce the computational complexity of computing the \emph{Mean F1 score} metric family ($F1a, F1h$ and $F1p$). Our technique is based on dedicated data structures described below. When loading the clusterings, we create a $rels$ hashmap relating the clusters to their elements (nodes) for each clustering. Besides the member nodes, our cluster data structure holds also:
\begin{inparaenum}[\itshape a\upshape)]
\item an accumulator $ctr$ for the matching contributions of the member nodes together with the pointer to the originating cluster 
from which these contributions are formed, and
\item the local contributions $cont$ for all members nodes of the cluster.
\end{inparaenum}
This data structure allows to evaluate the metrics using a single pass over all members of all clusters. The content of the $ctr$ attribute is reset on line~\ref{aln:ctrupd} in Algorithm~\ref{alg:mf1index} when adding a value of the contribution together with a distinct cluster pointer from the already stored one. 
The main procedure to evaluate the aforementioned
best matches for 
the clustering $cls$ is listed in Algorithm~\ref{alg:mf1index} considering
\begin{inparaenum}[\itshape a\upshape)]
\item the \texttt{fmatch} matching function (\texttt{f1} or \texttt{pprob} given in Eq.~\ref{eq:pprob}-\ref{eq:f1}) parameterized as $prob$ argument, and
\item the overlapping or multi-scale clusters evaluation semantics parameterized as $ovp$.
\end{inparaenum}
\begin{algorithm}[htbp]  
\caption{Best matches evaluation using efficient indexing technique}
\label{alg:mf1index}
\begin{algorithmic}[1]  
\Procedure{bmatches}{$grels, crels, cls, prob, ovp$}
\State $bms$ = vector()
\State fmatch = pprob if $prob$ else f1\label{aln:fmatch}    \Comment{Matching function}
\For{$c$ in $cls$}  \Comment{Traverse each cluster}
	\State $bmt$ = 0  \Comment{Value of the best match}
	\For{$nd$ in members($c$)}
		\State $ndcls$ = $grels$[$nd$]    \Comment{Node clusters}
		\For{$cn$ in $ndcls$}  \Comment{Evaluate node clusters}
			\State $cn.ctr$.add($c$, 1 / max(size($grels$[$nd$]), size($crels$[$nd$])) if $ovp$ else 1) \label{aln:ctrupd}
			\State $mt$ = fmatch($cn.ctr$, $cn.cont$, $c.cont$) 
			\If{$bmt < mt$}
				\State $bmt$ = $mt$
			\EndIf
		\EndFor
	\EndFor
	\If{$prob$}
		\State $bmt$ = $\sqrt{bmt}$
	\EndIf
	\State $bms$.append($bmt$)
\EndFor
\State \Return $bms$
\EndProcedure
\end{algorithmic}  
\end{algorithm}

\subsubsection{Evaluation and Constraints Matching}

Our new indexing technique reduces the computational complexity of Mean F1 Score evaluation from $O(N \cdot (|C| + |C'|))$~\cite{Chn14} $\simeq$\footref{ftn:clsnum} $O(N\sqrt{N})$ to $O(N \cdot s) \simeq O(N)$, where $N$ is the number of elements in the processing clusters and the constant $s$ is the average membership of the elements, which typically is $\in [1, 2)$ for overlapping clusterings. Implementations of all $MF1$ metrics ($F1a, F1h, F1p$) are provided in the open source \emph{\xms}\footref{ftn:xmsrc} utility and are available for free.
\al{TODO: Add evaluations figure on Large datasets and uncomment the sentence.}
The empirical satisfaction of the formal constraints for all $MF1$ metrics is given in Table~\ref{tbl:constrmf1} and discussed in Section~\ref{sec:discussion}.
\begin{table}[htbp]
\centering
\caption{Formal Constraints for MF1 metric family.}
\label{tbl:constrmf1}
\setlength\tabcolsep{2pt}  
\begin{tabular}{@{}l|cc|cc|cc|cc@{}}
\toprule
\multicolumn{1}{r|}{\textbf{\quad Clusterings}} & \multicolumn{2}{c|}{\textbf{Homogen.}} & \multicolumn{2}{c|}{\textbf{Complet.}} & \multicolumn{2}{c|}{\textbf{RagBag}} & \multicolumn{2}{c}{\textbf{SzQual.}} \\
\textbf{Metrics}                          & low                & high              & low                & high              & low              & high              & low               & high              \\ \midrule
F1a                    & 0.646              & 0.646             & \textit{0.639}              & \textit{0.663}             & 0.641              & 0.630               & \textit{0.795}             & \textit{0.936}             \\
F1h                    & 0.646              & 0.646             & \textit{0.639}              & \textit{0.660}             & 0.639              & 0.630               & \textit{0.795}             & \textit{0.935}             \\
\rowcolor{gray!30}
F1p                    & \textit{0.665}              & \textit{0.672}             & \textit{0.686}              & \textit{0.703}             & 0.693              & 0.693               & \textit{0.819}             & \textit{0.942}             \\ \bottomrule
\end{tabular}
\end{table}

\subsection{Generalized NMI}  
\label{sec:gnmi}

\subsubsection{Preliminaries}

Generalized NMI (GNMI)\footnote{\url{https://github.com/eXascaleInfolab/GenConvMI}\label{ftn:gecmi0}}~\cite{Esv12} uses a stochastic process to compare overlapping clusterings 
and feeds the random variables of the process into the standard definition of mutual information (MI)~\cite{Klb51}.
MI is evaluated by taking all pairs of clusters from the formed and ground-truth clusterings respectively and counts the number of common elements in each pair. 
Formally, given the ground-truth clustering $C'$ consisting of clusters $c' \in C'$ and the formed clusters $c \in C$, mutual information is defined as:
\begin{equation}
I(C':C) = \sum_{c' \ in C'} \sum_{c \ in C} p(c', c) \log_2{\frac{p(c',c)}{p(c')p(c)}}\;,
\label{eq:mi}
\end{equation}
where $p(c',c)$ is the normalized number of common elements in the pair of \textit{(category, cluster)}, $p(c')$ and $p(c)$ is the normalized number of elements in the categories and formed clusters respectively. The normalization is performed using the total number of elements in the clustering, i.e. the number of nodes in the input network.\\
Normalized Mutual information (NMI)~\cite{Dnn05} performs normalization of MI by \textit{maximum value, arithmetic or geometric mean} of the unconditional entropies $H(C')$ and $H(C)$ of the clusterings. 
Normalization by the maximum value of the entropies is the standard approach, which is also considered as the most discriminative one~\cite{Esv12,Mdd11}:
\begin{equation}
NMI(C',C) = \frac{I(C':C)}{\max{(H(C'), H(C))}}\;.
\label{eq:nmi}
\end{equation}
The unconditional entropy $H(X)$~\cite{Shn48} of clusters $x \in X$ is:
\begin{equation}
H(X) = -\sum_{x \in X}{p(x) \log_2 p(x)}.
\label{eq:entropy}
\end{equation}

\subsubsection{Proposed Extension and Optimizations}

The GNMI approach of using a stochastic process provides the only known way to evaluate \emph{standard NMI} for overlapping clusterings. The original GNMI implementation\footref{ftn:gecmi0} 
uses Intel's TBB library (lightweight treads) to execute the stochastic process on all available CPUs efficiently. 
However, the original implementation has several shortcomings, which makes it inapplicable to large datasets:
\begin{itemize}[leftmargin=*]
\item its hard-coded maximal number of stochastic events (successful samples) \verb|EVCOUNT_THRESHOLD| is not adequate for handling both small and large datasets. Moreover, the original value is too small for large datasets;
\item its fully random sampling of the cluster elements has a too high computational complexity on large datasets to produce results considering a reasonably small evaluation error (default value is 0.01);  
\item the two aforementioned points cause significant errors on small datasets and loss of convergence on large datasets while consuming significant computational resources.
\end{itemize}
We optimized and extended the original version addressing these issues 
in the following ways.
First, instead of the hard-coded \verb|EVCOUNT_THRESHOLD|, we dynamically evaluate the maximal number of stochastic events as:
\begin{equation}
evsmax = \max(\min(mbs', mbs), \frac{1}{rerr \sqrt{rrisk}}),
\label{eq:evsmax}
\end{equation}
where $mbs'$ and $mbs$ are the total membership of the elements in the categories and clusters respectively,
$rerr$ is the admissible error and $rrisk$ is the complement of the resulting confidence; $rerr$ and $rrisk$ are specified as input arguments with a default value equal to $0.01$.
\\
Second, we extended the original \texttt{try\_get\_sample} procedure with the weighed adaptive sampling 
given in Algorithm~\ref{alg:adapsamp}. 
The original version randomly takes the first node (cluster element) among all nodes in the clusterings as shown on line~\ref{aln:rnd1} and then
applies $mixer$ until it returns \texttt{false} or all nodes are traversed in a randomized order.
We traverse the nodes located in the same cluster of the formed ($c$) or ground-truth ($g$) clusterings on line~\ref{aln:optraver} and weight the shared nodes inversely to their membership. The weighting is performed to discount the contribution (importance) of frequent nodes compared to rare nodes since we traverse only a fraction of all nodes and they may have varying membership $\ge 1$. Indexes on the matched ground-truth and formed clusters are stored in the $mixer$ and their matching probability
is returned explicitly as $importance$.
\begin{algorithm}[htbp]  
\caption{Adaptive sampling of elements in GNMI}
\label{alg:adapsamp}
\begin{algorithmic}[1]  
\Procedure{try\_get\_sample}{$nodes, rrisk, mixer$}
\State $node$ = $nodes$[rand(size($nodes$))] \label{aln:rnd1}
\State $g, c$ = clsPair($node$)
\State $attempts$ = (size($g$) + size($c$)) / ($2 \cdot rrisk$)
\State $importance$ = 1 / max($\sqrt{\textnormal{size}(g) \cdot \textnormal{size}(c)}$, 1)
\State $adone$ = 1
\While{$mixer$.apply($g, c$) and ++$adone \le attempts$}\label{aln:optraver}
	\State $cm$ = rand($g, c$)
	\State $ndm$ = $cm$[rand(size($cm$))]
	\State $g, c$ = clsPair($ndm$)
	\State $importance$ += 1 / max($\sqrt{\textnormal{size}(g) \cdot \textnormal{size}(c)}$, 1)
\EndWhile
\State \Return $importance / adone$
\EndProcedure
\end{algorithmic}  
\end{algorithm}
\\
In addition, we performed some technical optimizations to reduce the number of memory allocations and copies, calls to external functions, etc.
Our \emph{extended implementation of GNMI}\footnote{\url{https://github.com/eXascaleInfolab/GenConvNMI}} is open source and available for free.

\subsubsection{Evaluation and Constraints Matching}

The empirical satisfaction of the formal constraints for $NMI$, the original $GNMI_{orig}$ and our $GNMI$ implementations is given in Table~\ref{tbl:constrnmi} and discussed below in Section~\ref{sec:discussion}.
\begin{table}[htbp]
\centering
\caption{Formal Constraints for NMI, original GNMI and our GNMI.}
\label{tbl:constrnmi}
\setlength\tabcolsep{2pt}  
\begin{tabular}{@{}l|cc|cc|cc|cc@{}}
\toprule
\multicolumn{1}{r|}{\textbf{\quad Clusterings}} & \multicolumn{2}{c|}{\textbf{Homogen.}} & \multicolumn{2}{c|}{\textbf{Complet.}} & \multicolumn{2}{c|}{\textbf{RagBag}} & \multicolumn{2}{c}{\textbf{SzQual.}} \\
\multicolumn{1}{l|}{\textbf{Metrics}}                          & low                & high              & low                & high              & low              & high              & low               & high              \\ \midrule
NMI                    & \textit{0.450}              & \textit{0.555}             & 0.546              & 0.546             & 0.434              & 0.434               & \textit{0.781}             & \textit{0.888}             \\
$\textnormal{GNMI}_{orig}$                    & 0.512              & 0.598             & 0.572              & 0.632             & 0.417              & 0.397               & 0.808             & 0.877             \\
\rowcolor{gray!30}
GNMI                    & \textit{0.448}              & \textit{0.557}             & 0.546              & 0.547             & 0.434              & 0.436               & \textit{0.781}             & \textit{0.888}             \\ \bottomrule
\end{tabular}
\end{table}
Since GNMI-s yields stochastic results, we report the median 
value over 5 runs with the same default values of the error and risk arguments.  
As the table clearly shows, our GNMI implementation is much more accurate than the original one and yields values equal to the original NMI within the specified admissible error (0.01) on non-overlapping clusterings. 
The empirical evaluation on
\begin{inparaenum}[\itshape a\upshape)]
\item the synthetic datasets formed using the LFR\footnote{\url{https://github.com/eXascaleInfolab/LFR-Benchmark_UndirWeightOvp}}~\cite{Lcn09b} framework and
\item the large real-world networks with ground-truth\footnote{\url{https://snap.stanford.edu/data/#communities}} introduced in~\cite{Yan15}
\end{inparaenum}
show that our implementation is one order of magnitude faster on datasets with $10^4$ nodes and two orders of magnitude faster on datasets with $10^6$ nodes than the original GNMI implementation. 
The actual computational complexity of both GNMI implementations depends on the structure of the clusters and on the number of overlaps in the clusterings. Moreover, for very dissimilar clusterings the evaluation might not converge at all, which is a disadvantage of 
the stochastic evaluation.  
The worst case computational complexity for our GNMI implementation is $O(N \cdot s \cdot |\bar{c}| \cdot |C|) \approx O(N \cdot |\bar{c}| \cdot |C|)$ while it is $O($\texttt{EVCOUNT\_THRESHOLD} $\cdot N \cdot |C|)$ for the original GNMI, where $N$ is the number of elements in the evaluating clusterings, $|C|$ is the number of clusters and $|\bar{c}|$ is the average size of the clusters.

\subsection{Discussion}
\label{sec:discussion}

When evaluating clusterings, it is important to 
\begin{inparaenum}[\itshape a\upshape)]
\item distinguish overlapping from multi-resolution (including hierarchical) clusterings and
\item consider the limitations of the applied accuracy metric
\end{inparaenum}
in order to produce meaningful results. First, we 
discuss the differences when handling overlapping clusterings versus multi-resolution clusterings having non-overlapping clusters on each resolution and how they affect accuracy evaluations.

In the case of overlapping clusterings, a node $x_i$ can be shared between $s$ clusters and has equal membership in each of them (e.g., a person may spend time for several distinct hobbies but the more hobbies are involved the less time the person devotes to each one). Thus, the membership contribution $x_i$ to each of the clusters is $1/s$. In case of non-overlapping, multi-resolution clusterings, a node $x_i$ may be a member of $s$ nested clusters taken at different resolutions (granularities), but here $x_i$ fully belongs to each of these 
clusters having a membership contribution equal to $1$ (e.g., a student tacking a course can be a full member of the course, as well as of the department offering the course and of the whole university).
These distinct semantics for the elements shared in a clustering are represented by the $ovp$ argument in our \emph{\xms}\footref{ftn:xmsrc} utility for all MF1 metrics.

We summarize the advantages and limitations of the various metrics discussed in this paper as follows:
\\
\textbf{Omega Index pros:} its values are not affected by the number of clusters (unlike NMI) and have an intuitive interpretation (0 means equal quality to a random clustering).
\\
\textbf{Omega Index cons:} it performs purely 
when applied to multi-resolution clusterings 
and has the highest computational complexity among all considered metrics ($O(N^2)$).
\\
\textbf{MF1 pros:} it has the lowest computational complexity (linear when using our indexing technique).
\\
\textbf{MF1 cons:} it evaluates the best-matching clusters only, which 
gives an unfair advantage to the clusterings with larger numbers of clusters (which is partially addressed by the application of the harmonic mean in $F1h$ and $F1p$ instead of the average).
\\
\textbf{GNMI pros:} it parallelizes very well and inherits NMI's pros, namely it evaluates the full matching of all clusters (unlike MF1). Also, it is well-grounded formally in Information Theory.
\\
\textbf{GNMI cons:} the convergence in the stochastic implementation is not guaranteed (though loss of convergence typically indicates a relevance close to zero); the stochastic process yields non-deterministic results and is computationally heavy; its execution time is hard to estimate. In addition, GNMI inherits the NMI's cons, namely the results depends on the number of clusters being evaluated and increase up to $\approx 0.3$ 
for large numbers of clusters.

\begin{table}[htbp]
\centering
\caption{Formal Constraints for Omega Index, MF1 and GNMI.}
\label{tbl:constr}
\begin{tabular}{@{}lcccc@{}}
\toprule
\multicolumn{1}{r}{\textbf{Metrics\textbackslash{}Clusterings}} & \textbf{Homogen.} & \textbf{Complet.} & \textbf{RagBag} & \textbf{SzQual.} \\ \midrule
{[}Soft{]} Omega Index                                          & +                 & +                 &                 &                  \\
F1h                                                             &                   & +                 &                 & +                \\
F1p                                                             & +                 & +                 &                 & +                \\
GNMI                                                            & +                 &                   &                 & +                \\ \bottomrule
\end{tabular}
\end{table}
Empirical evaluation of the formal constraints satisfaction for all discussed metrics on the original intuitive samples from~\cite{Amg09} is given in Table~\ref{tbl:constr}. As it shown in the table, none of the metrics satisfies the RagBag constraint, which is essential in practice when evaluating the multi-resolution or hierarchical clusterings since the more fine-grained (higher resolutions) are expected to have less noisy structure. Otherwise, the proposed $F1p$ metric performs the best according to the empirical satisfaction of the formal constraints having the lowest computational complexity
(linear on the number of elements being clustered)
compared to GNMI and Omega Index. 

%
%
%
%
%
%
%
%
%
%
%
%
%
%
%

\section{Conclusions}  
\label{sec:conclusion}

In this paper, we discussed the state-of-the-art accuracy metrics applicable to overlapping clustering evaluations on large datasets and introduced several optimizations and extensions of the discussed metrics. In particular, we introduced an efficient indexing technique to speedup the evaluation of Mean F1 score from $O(N\sqrt{N})$ to $O(N)$, where $N$ is the number of elements in the clustering. We proposed an adaptive sampling strategy for \emph{GNMI}, which not only speeds up the evaluation by orders of magnitude but also improves the precision of the metric. In addition, we proposed two extensions of the Average F1 score ($F1a$), namely $F1h$ and $F1p$. $F1h$ addresses the issues of the loss of indicativity of $F1a$ in the range $[0, 0.5]$ while $F1p$ empirically improves the satisfaction of the formal constraints we consider
without sacrificing efficiency. We also proposed an extension of the Omega Index called \emph{Soft Omega Index}, which is equivalent to the original Omega Index evaluating non-overlapping clusterings and yields more discriminative results for overlapping clusterings due to the fact that it considers partial matches of pairs of elements.

Besides the proposed optimizations and extensions of the accuracy metrics, we discussed formal constraints for each metric, as well as their applicability and limitations for specific cases in overlapping and multi-resolutions clusterings. Our analysis of the metrics should provide insight for their future usage and should hopefully help identify the best performing clustering algorithm for particular user's needs or tasks.

We freely provide implementations of all discussed metrics, and are also integrating them into an open source benchmarking framework for clustering algorithms evaluation, \bmark\unskip\footref{ftn:bmsrc}.

\bibliographystyle{IEEEtran}\balance
\bibliography{IEEEabrv,../xmeasures}

\end{document}